\documentclass[a4paper, 10pt]{article}

\newcommand{\iTitle}[1]{\begin{center}\Large\bf #1\end{center}}
\newcommand{\iAuthor}[1]{\begin{center}\small #1\end{center}}
\newcommand{\iAddress}[1]{\begin{center}\small\it #1\end{center}}
\newcommand{\iAbstract}[1]{\noindent {\large\bf Abstract}\\#1}
\newcommand{\iKeywords}[1]{\noindent{\it Keywords:} #1 }

\usepackage{multirow}
\usepackage{geometry}
\usepackage{indentfirst} 
\usepackage[pagewise,switch]{lineno}
\usepackage{graphicx}
\usepackage{subfigure}
\usepackage{color}
\usepackage{xcolor}
\definecolor{red}{rgb}{1,0,0}
\usepackage[colorlinks=true, linkcolor=blue, urlcolor=magenta, citecolor=red]{hyperref}
\newcommand{\refb}[1]{(\ref{#1})}
\newcommand{\iRef}[7]{#1, {\it``#2"}, \href{#7}{#3, {\bf#4}, #5(#6).} }

\newcommand{\draftnote}[1]{{#1}}

\newcommand{\bw}{\begin{widetext}}
\newcommand{\ew}{\end{widetext}}
\newcommand{\be}{\begin{equation}}
\newcommand{\en}{\end{equation}}
\newcommand{\bee}{\begin{equation}}
\newcommand{\ene}{\end{equation}}
\newcommand{\bea}{\begin{eqnarray}}
\newcommand{\ena}{\end{eqnarray}}

\newcommand{\eqref}[1]{Eq.~(\ref{#1})}

\def\pslash{p\!\!\!\slash }
\def\Aslash{A\!\!\!\slash }

\def\to{\rightarrow}

\def\pslash{p\!\!\!\slash }

\def\ie{{\it i.e.}}

\begin{document}

\iTitle{Time-dependent Aharonov-Bohm effect on the noncommutative space}

\vspace{0.3cm}
\iAuthor{Kai Ma\footnote{\href{mailto:makainca@yeah.net}{makainca@yeah.net}}, Jian-Hua Wang}
\vspace{-0.6cm}
\iAddress{Department of Physics, Shaanxi University  of Technology, Hanzhong, 723001, Peoples Republic of China}

\iAuthor{Huan-Xiong Yang}
\vspace{-0.6cm}
\iAddress{Interdisciplinary Center for Theoretical Study, University of Science and Technology of China, Hefei 200026, Peoples Republic of China}

\vspace{0.5cm}
\noindent\rule[0.25\baselineskip]{\textwidth}{0.8pt}
\iAbstract{
We study the time-dependent Aharonov-Bohm effect on the noncommutative space. Because there is no net Aharonov-Bohm phase shift in the time-dependent case on the commutative space, therefore, a tiny deviation from zero indicates new physics. Based on the Seiberg-Witten map we obtain the gauge invariant and Lorentz covariant Aharonov-Bohm phase shift in general case on noncommutative space. We find there are two kinds of contribution: momentum-dependent and momentum-independent corrections. For the momentum-dependent correction, there is a cancellation between the magnetic and electric phase shifts, just like the case on the commutative space. However, there is a non-trivial contribution in the momentum-independent correction. This is true for both the time-independent and time-dependent Aharonov-Bohm effects on the noncommutative space. However, for the time-dependent Aharonov-Bohm effect, there is no overwhelming background which exists in the time-independent Aharonov-Bohm effect on both commutative and noncommutative space. Therefore, the time-dependent Aharonov-Bohm can be sensitive to the spatial noncommutativity. \draftnote{The net correction is proportional to the product of the magnetic fluxes through the fundamental area represented by the noncommutative parameter $\theta$, and through the surface enclosed by the trajectory of charged particle.} More interestingly, there is an anti-collinear relation between the logarithms of the magnetic field $B$ and the averaged flux $\Phi/N$ (N is the number of fringes shifted). This nontrivial relation can also provide a way to test the spatial noncommutativity. For $B\Phi/N\sim 1$, our estimation on the experimental sensitivity shows that it can reach the $\rm 10GeV$ scale. This sensitivity can be enhanced by using stronger magnetic field strength, larger magnetic flux, as well as higher experimental precision on the phase shift.
}

\vspace{0.3cm}

\iKeywords{Noncommutative geometry, Aharonov-Bohm effect, Geometry phase}

\noindent\rule[0.25\baselineskip]{\textwidth}{0.8pt}

\tableofcontents
\newpage

\section{Introduction}\label{intro}
The proposal of a discrete space-time with noncommutative algebra was motivated by the infinities in the Quantum Field Theory (QFT)~\cite{1947QuantizedSpaceTime-Snyder}, and represented by the the commutation relation between the canonical position operators as follows,
\bee\label{eq:ncdefine}
[x_{\mu}, x_{\nu}] = i \theta_{\mu\nu}\,,
\ene
where $\theta_{\mu\nu}$ is a totally anti-symmetric constant tensor representing the strength and relative directions of the noncommutativity, and has dimension of length-square. The nontrivial commutation relation implies that we have an intrinsic uncertainty in the position space, and hence the physical properties of the quantum system are dramatically changed. Even though the initial proposal is purely theoretical, it was shown in later that the commutation relation \refb{eq:ncdefine} can appear in the dynamics of charged particles in the presence of electromagnetic fields~\cite{Peierls:1933,Gamboa:2001,Ho:2002,Ma:2011,Wang:2013,Wang:2015}. Particularly, it was also shown that the noncommutative algebra can emerge from string theory embedded in a magnetic background~\cite{SW:1999}, as well as the quantum gravity~\cite{Freidel:2006,MirFaizal:2013:A,Gangopadhyay:2016}. Extensions with non-anticommutative algebra were also studied~\cite{MirFaizal:2013:B,MirFaizal:2013:C,Vasyuta:2016}. As a fundamental theory, the noncommutativety of space is introduced through the Moyal-Weyl product (also called $\star$-product),
\begin{equation}\label{SP}
f( x ) \star g( x )
= \exp{\left[\frac{i}{2} \theta_{\mu\nu} \partial_{x_{\mu}}\partial_{y_{\nu}}\right]} f(x)g(y)|_{x=y},
\end{equation}
where $f(x)$ and $g(x)$ are two arbitrary infinitely differentiable functions on the commutative $R^{3+1}$ space-time. Based on the $\star$-product, the gauge theory can be established in the usual way. At the leading order of noncommutative parameter $\theta_{\mu\nu}$, the noncommutative correction is represented by the transformation $x^{\mu} \to x^{\mu} + \theta^{\mu\nu}p_{\nu}/2$, which is called Bopp's shift \cite{TCurtright:1998}. 

The noncommutative extension of gauge field theory possesses many richer distinguished physical properties. It was shown that while the translational invariance is preserved, the Lorentz symmetry is violated~\cite{Douglas:2001,Szabo:2003}. Because of this, the degeneracy of the energy levels of hydrogen atom is removed~\cite{Chaichain:2001}. This can also affect the motion of dipole in the electromagnetic fields~\cite{Zhang:2004}. Particularly, the topological properties of the ordinary gauge field theory, for instances, the Aharonov-Bohm (AB) effect~\cite{AB:1959} and the Aharonov-Casher (AC) effect~\cite{AC:1984}, can also receive corrections. \draftnote{Several approaches have been used to study the noncommutative corrections on AB effect. In Refs.~\cite{Chaichian:2001B,Chaichian:2002}, the path integral method was used, and the results satisfy the noncommutative gauge symmetry, which is a shortage of the Bopp's shift method used in Refs.~\cite{Li:2006,Mirza:2004,Li:2007,Wang:2007}. In Refs.~\cite{Harms:2007,Gangopadhyay:2015}, a linear transformation between the noncommutative and commutative canonical position and momentum operators was employed to obtain the noncommutative corrections. Similar method was also used in Ref.~\cite{Bertolami:2005} for the gravitational quantum well system, and in Ref.~\cite{Liang:2015} for the quantum ring. However, except the appearance of momentum-momentum noncommutativity which can not break the original electromagnetic gauge symmetry\cite{Bertolami:2015}, the solution of such linear transformation is exactly the same as the one obtained by using the Bopp's shift method. On the other hand, it was pointed out that, the naive path integral formulation of the noncommutative quantum mechanics and the Bopp's shift approach lead neither to a gauge invariant nor to a gauge covariant AB phase factor~\cite{Bertolami:2015,Chaichian:2008}, and the noncommutative extension of Wilson line method can solve this problem~\cite{Chaichian:2008}. Recently, the gauge symmetry in the noncommutative Proca model was also studied in Ref.~\cite{Abreu:2016}.

On the other hand, the more general solution to this gauge symmetry problem is provided by the Seiberg-Witten (SW) map \cite{SW:1999} from the noncommmutative space to the ordinary one \cite{Brace:2001, Barnich:2001}. By using the SW map, Carroll {\it et. al.} shown that \cite{Carroll:2011}, the noncommutative model is a subset of a general Lorentz-violating standard-model extension~\cite{Colladay:1997,Colladay:1998}, and the noncommutative parameter $\theta$ is strongly constrained by the clock-comparison tests, $\sqrt{1/\theta} > 10\rm{TeV}$~\cite{Carroll:2011}. It has also been used to study the axial anomaly~\cite{Banerjee:2002}, Chern-Simons~\cite{Picariello:2002,SubirGhosh:2005}, Galilean symmetry~\cite{Chakraborty:2004}, Schwinger Model~\cite{Saha:2006}, as well as the noncommutative 1-cocycle~\cite{Jackiw:2002}. Extension involving spin degree of freedom was proposed in Ref.~\cite{Ma:2016}. The $\theta$-exact SW map was also studied~\cite{Fidanza:2002,WWang:2011,Martin:2012aw}, and have been used to analyze the noncommutative effects in high energy collision processes~\cite{WWang:2012,WWang:2013}.

In this paper, we study the noncommutative corrections on the time-dependent AB effect~\cite{Rousseaux:2008,Moulopoulos:2010,Singleton:2013,Singleton:2014,Singleton:2015}. It was shown that because of the gauge and Lorentz symmetries, the time-dependent AB phase shift vanishes on the commutative space. This is true even for the time-dependent non-Ablian AB effect~\cite{Mansoori:2016}. Therefore, a tiny derivation from zero in the phase shift indicates new physics. While the ordinary electromagnetic gauge symmetry have to be kept in order to get consistent physical results, the Lorentz symmetry is broken in the noncommutatively extended models. Therefore, it is expected that the time-dependent AB system can provide an ideal tool to measure the noncommutative parameter $\theta^{\mu\nu}$. Because both gauge and Lorentz symmetries play essential roles in the time-dependent AB effect, hence, it is necessary to have a Lorentz covariant and gauge invariant formulation of the noncommutative corrections. As we have mentioned, the SW map is a general method to preserve the ordinary gauge symmetry on noncommutative space, and it is naturally formulated in a Lorentz covariant way. Therefore, in this paper, we will use the SW map to study the noncommutative corrections on the time-dependent AB effect.
}

The contents of this paper are organized as follows: in Sec.~\ref{sec:nccovariantAB} we study the covariant formulation of the AB phase shift in general case on the noncommutative space; in Sec.~\ref{sec:ncexpAB} we will study the noncommutative corrections on both time-independent and time-dependent AB phase shifts; in Sec.~\ref{sec:exp}, the experimental schemes of detecting the spatial noncommutativity are studied based on the time-dependent AB effect, and the experimental sensitivity is also estimated; our conclusions are given in the final section, Sec. \ref{conclusion}.

\section{Covariant nonvommutative AB effect}\label{sec:nccovariantAB}
The AB effect~\cite{AB:1959} is one of the most profound physical property of the quantum gauge theory. It is predicted that the quantum phase of charged particle wave can be shifted by the pure electromagnetic gauge potential $A_{\mu}$ without local interactions with the electromagnetic field strength $\vec{B}$ and $\vec{E}$. There are two components in the AB phase, magnetic and electric effects, corresponding to the vector potential $\vec{A}$ and the scalar potential $\varphi$, respectively. It has been applied in various quantum systems, and has been employed to explore the physical effects of the noncommutative space. In this section we will study the general AB phase on the noncommutative space. To this end, it is necessary to formulate the magnetic and electric AB phase shifts in a covariant way. Various approaches has been studied on the ordinary space~\cite{Rousseaux:2008,Moulopoulos:2010,Singleton:2013,Singleton:2015}. Here we study its noncommutative corrections by starting from the fundamental Lagrangian. For spin-$1/2$ particle with charge $Q$ interacting with the electromagnetic fields, the Lagrangian is, 
\bee\label{eq:lagrangian}
\mathcal{L} =
\bar{\psi}(x)( \pslash - Q \Aslash - m ) \psi(x) \,.
\ene
From it we can obtain the corresponding equation of motion of the charged particle. The formal solution is $\psi(x) = e^{i\phi(x)}\psi_0(x)$, where $\psi_0(x)$ is the free solution, and the phase factor
\bee\label{ab}
\phi(x) = \int_{x_0}^{x} A_{\mu}(z) dz^{\mu} 
%= \frac{1}{2}\int_{x_0}^{x} F_{\mu\nu}(z)dS^{\mu\nu} \,.
\ene
The AB phase shift corresponds to the case of closed path, and the magnetic contribution corresponds to the case with $\vec{E}=0$, the electric AB phase corresponds to $\vec{B}=0$~\cite{AB:1959}. %Note that the electromagnetic field $\vec{B}$ and $\vec{E}$ vanish only in the region of matter particle moving. This is important to get nontrivial phase factor~\cite{AB:1959}.

On the noncommutative space, the Lagrangian \refb{eq:lagrangian} is deformed by the noncommutative algebra in \eqref{eq:ncdefine}, and can be written as follows,
\bee\label{nclag}
\mathcal{L} =
\bar{\psi}(x)( \pslash - Q \Aslash - m ) \star \psi(x) \,.
\ene
The corresponding noncommutative gauge transformations are~\cite{Chaichian:2001B,Chaichian:2002,Chaichian:2008}
\bea
\label{gauge-transformation-matter}
\psi'(x) &=& U(x)\star \psi(x) \,,\\
\label{gauge-transformation-gauge}
A'_{\mu}(x) &=& U(x)\star \bigg( A_{\mu}(x) -\frac{ 1 }{ Q } p_{\mu} \bigg) \star U^{-1}(x)\,.
\ena
The topological AB phase related to this noncommutative gauge symmetry has been investigated~\cite{Chaichian:2001B,Chaichian:2002,Li:2006,Harms:2007,Gangopadhyay:2015,Chaichian:2008}. However, it was pointed out the naive path integral formulation of the noncommutative quantum mechanics and the Bopp's shift approach lead neither to a gauge invariant nor to a gauge covariant AB phase factor~\cite{Chaichian:2008}. Even though it was shown that, the extension of Wilson line method can be used to define a gauge invariant AB phase~\cite{Chaichian:2008}, but this method can not be directly used for studying of noncommutative effects in other quantum system. The more general solution to this gauge symmetry problem is provided by the SW map \cite{SW:1999} from the noncommmutative space to the ordinary one \cite{Brace:2001, Barnich:2001}. Furthermore, it is naturally formulated in a Lorentz covariant way. Hence, it was used in various quantum system, ranging form low energy atom physics~\cite{Carroll:2011} to the hight energy collision processes~\cite{WWang:2012,WWang:2012,WWang:2013}. We will use the SW map to study the noncommutative corrections. For the $U(1)$ gauge symmetry the SW maps are given by
\bea
\psi &\to & \psi - \frac{1}{2} Q \theta^{\alpha\beta} A_{\alpha} \partial_{\beta}\psi\,,  \\
A_{\mu} &\to & A_{\mu} - \frac{1}{2} Q\theta^{\alpha\beta} A_{\alpha} ( \partial_{\beta} A_{\mu} + F_{\beta\mu} )\,,
\ena
where $F_{\mu\nu} = \partial_{\mu} A_{\nu} -  \partial_{\nu} A_{\mu} $ is the electromagnetic field strength tensor. Then in terms of the ordinary fields the noncommutative Lagrangian \refb{nclag} can be written as,
\bee\label{ac-nc}
\mathcal{L}_{NC} =
\big( 1 - \frac{1}{4}Q\theta^{\alpha\beta} F_{\alpha\beta} \big) \bar{\psi}(x) ( i\gamma_{\mu} \mathcal{D}^{\mu}  - m )  \psi(x) +
\frac{i}{2} Q\theta^{\alpha\beta} \bar{\psi}(x) \gamma^{\mu} F_{\mu\alpha} \mathcal{D}_{\beta} \psi(x) \,,
\ene
where $\mathcal{D}_{\beta} =  \partial_{\beta} + i Q A_{\beta}$ is the covariant derivative. Because this Lagrangian involves only the covariant derivative $\mathcal{D}_{\beta}$ and the electromagnetic field strength $F_{\mu\nu}$, therefore, it is gauge invariant under the $U(1)$ gauge symmetry. In this sense, the noncommutative corrections on the AB phase shift can be defined unambiguously, and can be interpreted consistently on the commutative and noncommutative spaces.

\draftnote{The Lagrangian \refb{ac-nc} contains two kinds of corrections. The first term, proportional to $\big( 1 - Q\theta^{\alpha\beta} F_{\alpha\beta}/4 \big)$, comes from the coupling between the background field $\theta_{\mu\nu}$ and the electromagnetic field strength $F_{\mu\nu}$. Because the AB phase shift is related to the covariant derivative, therefore, it is expected that the first kind of correction does not affect the AB effect. In an exact treatment, this term can give a correction on the charge $Q$ of the matter particle~\cite{Carroll:2011}, which can certainly affect the AB phase in turn. However, the correction turns out to be second order of the noncommutative parameter $\theta_{\mu\nu}$, hence can be neglected consistently. This property can be realized by investigating the equation of motion which can be obtained from the Lagrangian \eqref{ac-nc} as follows,
\bee\label{ncmotioneq}
 ( i\gamma_{\mu} \mathcal{D}^{\mu}  - m )  \psi(x) +
\frac{i}{2} Q\big( 1 - \frac{1}{4}Q \theta^{\alpha\beta} F_{\alpha\beta} \big) ^{-1} \theta^{\alpha\beta}\gamma^{\mu} F_{\mu\alpha} \mathcal{D}_{\beta} \psi(x) =0\,.
\ene
Here we have multiplied a factor $\big( 1 - \frac{1}{4}Q \theta^{\alpha\beta} F_{\alpha\beta} \big) ^{-1}$ to normalize the kinematical energy. By expanding $\big( 1 - \frac{1}{4}Q \theta^{\alpha\beta} F_{\alpha\beta} \big) ^{-1}$ with respect to the noncommutative parameter $\theta^{\mu\nu}$, one can see that the first kind of correction can be neglected at the first oder of the noncommutative parameter $\theta^{\mu\nu}$. Therefore, from here and after, we will focus on the second term which involves the correction on the covariant derivative $\mathcal{D}_{\beta}$.}

The second correction proportional to $\theta^{\alpha\beta} F_{\mu\alpha} \mathcal{D}_{\beta}$, comes from the coupling between the background field $\theta_{\mu\nu}$,  the electromagnetic field strength $F_{\mu\nu}$, and the covariant derivative $\mathcal{D}_{\beta}$. Therefore, it is expected that the AB phase shift will receive non-trivial corrections. By neglecting the first kind of noncommutative correction, the equation of motion \refb{ncmotioneq} can be written as
\bea\label{ncacfinald}
&& ( i\gamma_{\mu} \mathcal{D}_{NC}^{\mu}  - m ) \psi(x) = 0 \,, 
\\[2mm]
\label{ncacfinald2}
&& \mathcal{D}_{NC}^{\mu}  = \big( g^{\mu}_{~~\beta} + \frac{1}{2} Q F^{\mu\alpha}\theta_{\alpha\beta} \big) \mathcal{D}^{\beta}\,.
\ena
This result is different from the previous results obtained by using the Bopp's shift method. If we use the Bopp's shift method, then the noncommutative corrected covariant derivative is $\mathcal{D}_{NC}^{\mu} =  \big( g^{\mu\beta} + \frac{i}{2} Q\theta^{\alpha\beta} \partial_{\alpha} A^{\mu} (x) \big)\mathcal{D}_{\beta}$, which is obviously not gauge invariant under the $U(1)$ gauge transformation. Therefore, noncommutative corrections can not be defined unambiguously. However, the correction in \eqref{ncacfinald2} is proportional to the covariant derivative, and depends only on the electromagnetic field strength $F^{\mu\alpha}$. Therefore, it is invariant under the ordinary electromagnetic $U(1)$ gauge transformation.

The AB phase can be obtained by solving the equation of motion, \eqref{ncacfinald}. At the leading order of the noncommutative parameter, the AB phase shift can be written as
\bee
\phi_{NC}^{AB} = \phi^{AB} + \phi_{\theta-v}^{AB} + \phi_{\theta-g}^{AB}\,.
\ene
The first term $\phi^{AB}$ is the ordinary AB phase. The second term $\phi_{\theta-v}^{AB}$ is a  momentum-dependent noncommutative correction,
\bee\label{ncabp}
\phi_{\theta-v}^{AB} 
%=  \oint C^{\mu\nu} p^{\nu} d x_{\mu}
=  \frac{Q}{2} \oint  F^{\mu\alpha} \theta_{\alpha\beta}  p^{\beta} d x_{\mu}
%=  \frac{Q}{2} \theta_{\alpha\beta}  p^{\beta}\oint  \partial^{\nu}F^{\mu\alpha} d S_{\mu\nu}
%=  \frac{Q}{4} \theta_{\alpha\beta}  p^{\beta}\oint  \partial^{\alpha}F^{\mu\nu} d S_{\mu\nu}
\,.
\ene
This represents the general non-local property of the electromagnetic interactions on the noncommutative space. The third term $\phi_{\theta-g}^{AB}$ is a momentum-independent noncommutative correction,
\bee\label{ncaba}
\phi_{\theta-g}^{AB} 
=  -\frac{Q^2}{2} \oint  F^{\mu\alpha} \theta_{\alpha\beta}  A^{\beta} d x_{\mu}
%=  -\frac{Q^2}{4} \theta_{\alpha\beta}  A^{\beta}\oint  \partial^{\alpha}F^{\mu\nu} d S_{\mu\nu}
%- \frac{Q^2}{2} \oint  F^{\mu\alpha} \theta_{\alpha\beta}  \partial^{\nu} A^{\beta} d S_{\mu\nu}
\,.
\ene
This term appears due to the gauge covariance of the noncommutative correction in \eqref{ncacfinald2}. Above results are the gauge invariant and Lorentz covariant AB phase shift on the noncommutative space. It is worthy to note that, unlike the ordinary AB phase shift in \eqref{ab}, both the momentum-dependent and momentum-independent phase shifts in \eqref{ncabp} and \eqref{ncaba} involve the local interactions between the charged particle and electromagnetic field strength $F^{\mu\nu}(x)$. This property is very important to receive non-trivial noncommutative correction on the time-dependent AB effect, the details will be discussed in next section.

%%%%%%%%%%%%%%%%%%%%%%%%%%%%%%%%%%%%%%%%%%%%%
\section{Noncommutative time-dependent AB effect}\label{sec:ncexpAB}
On the ordinary space, it has been shown that there is an exact cancellation between the magnetic and electric AB phase shifts, and therefore, the net phase shift vanishes~\cite{Rousseaux:2008,Moulopoulos:2010,Singleton:2013,Singleton:2015}. The cancellation happens because the phase shift induced by the magnetic field $\vec{B}(t) = \vec{\nabla}\times\vec{A}(t)$ is the negative of the one induced by the electric field $\vec{E}(t) = - \partial_{t}\vec{A}(t)$ (in the gauge of vanishing electric potential, $\phi=0$). However, this can not be true on the noncommutative space, because the noncommutative corrections involve the local interactions between the charged particle and the electromagnetic field strength, and also violate the Lorentz symmetry. In this section we study the physical properties of the noncommutative corrections on the time-dependent AB effect. 

Let us specify the physical configuration in first. We use the standard AB configuration, and consider an infinite solenoid with electric current $I(t)$ which creates a vector potential outside the solenoid as follows
\bee\label{afield}
\vec{A}(t, \vec{x}) = \frac{k I(t)}{r} \vec{e}_{\phi}\,,
\ene   
where $k$ is a constant whose exact form is not important for now and $\vec{e}_{\phi}$ is the unite vector in the azimuthal angle direction in the $x-y$ plane. For convenience and generality, we have show the time-dependence of the electric current explicitly. The time-independent AB effect corresponds to the case with $\partial_{t}I(t)=0$. In this configuration, the magnetic field $\vec{B}(t) = \vec{\nabla}\times\vec{A}(t)$ is along the $z$-direction, and the electric field $\vec{E}(t) = - \partial_{t}\vec{A}(t)$ lies in the $x-y$ plane. Therefore, the electromagnetic field strength tensor is
\bee
F^{\mu\nu} 
=
\left(\begin{array}{cccc}
0 & -E^{1} & -E^{2} & 0 \\
E^{1} & 0 & -B^{3} & 0\\
E^{2} & B^{3} & 0 & 0 \\
0 & 0 & 0 & 0
\end{array}\right)\,.
\ene 
Without loss of generality, we can further assume the motion of charged particle is comfined in the $x-y$ plane, \ie, $v_z \approx 0$. For completeness we also give our conventions on the noncommutative parameter $\theta^{\mu\nu}$ as follows,
\bee
\theta^{\mu\nu}
=
\left(\begin{array}{cccc}
0 & 0 & 0 & 0 \\
0 & 0 & \theta^{12} & \theta^{13}\\
0 & -\theta^{12} & 0 & \theta^{23} \\
0 & -\theta^{13} & -\theta^{23} & 0
\end{array}\right)
=
\left(\begin{array}{cccc}
0 & 0 & 0 & 0 \\
0 & 0 & \theta^{3} & -\theta^{2}\\
0 & -\theta^{3} & 0 & \theta^{1} \\
0 & \theta^{2} & -\theta^{1} & 0
\end{array}\right)
\,,
\ene 
where we have defined new parameters by the relation $\theta^{i}=\epsilon^{ijk}\theta^{jk}/2$. Furthermore, we have resricted our in the case of $\theta^{0i}=0$, because the time-position noncommutativity can violate the unitarity~\cite{Chaichian:2001B,Chaichian:2002,Chaichian:2008,Li:2006,Liang:2015}, which beyonds the contents of this paper.

Before we go on to study the time-dependent AB phase shift, it is necessary to disucs the time-independent AB phase shift in first. As we have mentioned, a general property of both the momentum-dependent and momentum-independent noncommutative corrections is that the corrections involve the local interactions between the charged particle and electromagnetic field strength $F^{\mu\nu}(x)$, see \eqref{ncabp} and \eqref{ncaba}. However, for the momentum-dependent correction \refb{ncabp}, there is no singularity, and the electromagnetic field strength $F^{\mu\nu}(x)$ vanishes at the location of charged particle, therefore, there is no net contribution. This property can also be understood by using the Stokes's theorem by which the momentum-dependent correction can be written as
\bee\label{ncabps}
\phi_{\theta-v}^{AB} 
=  \frac{Q}{4} \theta_{\alpha\beta}  p^{\beta}\oint  \partial^{\alpha}F^{\mu\nu} d S_{\mu\nu}\,.
\ene
Because the electromagnetic field strength $F^{\mu\nu}$ is a constant over the whole space-time for time-independent AB effect, therefore, $\partial^{\alpha}F^{\mu\nu} =0$ and hence there is no non-trivial phase shift. However, this is not true for the momentum-independent correction in \eqref{ncaba}. This is because there is a singularity in the vector potential $A_{\mu}$. By using the Stokes's theorem the momentum-independent correction can be written as,
\bee\label{ncabas}
\phi_{\theta-g}^{AB} 
=  -\frac{Q^2}{4} \theta_{\alpha\beta}  \oint A^{\beta} \partial^{\alpha}F^{\mu\nu} d S_{\mu\nu}
- \frac{Q^2}{2} \oint  F^{\mu\alpha} \theta_{\alpha\beta}  \partial^{\nu} A^{\beta} d S_{\mu\nu}\,.
\ene
The first term vanishes because of $\partial^{\alpha}F^{\mu\nu} =0$. However, the second term can give non-trivial contribution. For the magnetic component we have,
%Let us first consider the momentum-dependent noncommutative corrections $\widetilde{\phi}_{\theta-v}^{AB}$, the magnetic part is 
\bee\label{ncabam}
\phi_{\theta-g}^{AB-M} 
=  -\frac{Q^2}{2} \oint  F^{ij} \theta_{jk}  A^{k} d x_{i}
= -\frac{Q^2}{2} \oint  \big( \vec{B}\cdot\vec{A} \big)  \vec{\theta}\cdot d\vec{x}
 +\frac{Q^2}{2} \oint \big( \vec{B}\cdot \vec{\theta} \big)  \vec{A} \cdot d\vec{x}
 =  \frac{Q}{2} \big( \vec{B}\cdot \vec{\theta} \big) \phi^{AB} \,,
\ene
where we have used a relation $\vec{B}\perp\vec{A}$, \ie, $\vec{B}\cdot\vec{A}=0$, which is a result of the physical configuration \refb{afield}, and is necessary to create the singularity of the space. Therefore, there is a nontrivial contribution on the magnetic AB phase shift which is proportional to the magnetic flux through the fundamental area represented by the noncommutative parameter $\vec{\theta}$. The electric AB phase shift is
\bee
\phi_{\theta-g}^{AB-E} 
=  -\frac{Q^2}{2} \oint  F^{0j} \theta_{jk}  A^{k} d t
= -\frac{Q^2}{2} \oint   \vec{\theta}\cdot (\vec{E}\times\vec{A}) d t
= 0 \,,
\ene
where we have used the static condition $\vec{E} = \partial_{t}\vec{A}=0$ for the time-independent AB phase. Therefore, the net phase shift for the static AB effect can be written as
\bee\label{netncab}
\phi_{NC}^{AB} = \bigg(1 + \frac{1}{2} Q \vec{B}\cdot \vec{\theta} ~\bigg)\phi^{AB}\,,
\ene
which scales the ordinary AB phase shift by a factor of $1 +  Q \vec{B}\cdot \vec{\theta}/2$. In the consideration of that the noncommutative property of space happens at the Plank scale, it is very hard to create such strong magnetic field so that $\vec{B}\cdot \vec{\theta} \sim 1$ to measure the spatial noncommutativity. In Ref.~\cite{Chaichian:2002}, a lower bound $\sqrt{\theta^{-1}}>10^{-6}\rm{GeV}$ was obtained by using the time-independent AB effect.

Let us go on to study the time-dependent AB phase shift on the noncommutative space. In this case the momentum-dependent correction can be non-zero, because the electromagnetic field does not vanish at the location of the charged particle wave. However, we will show that this kind of correction still vanishes because of the gauge invariance. For clarity, from here and after, we will add a tilde ``$~\widetilde{}~$" on the phase shifts in the time-dependent case to distinguish from the time-independent case. The magnetic and electric phase shift of $\phi_{\theta-v}^{AB}$ are
\bee\label{eq:ncabpt}
\widetilde{\phi}_{\theta-v}^{AB-M} 
%=  \frac{1}{4} Q m \theta_{\alpha\beta}v^{\beta}\oint  \partial^{\alpha}F^{ij} d x_{i} \wedge dx_{j}\,.
=  \frac{1}{2} Q m \theta_{\alpha\beta}v^{\beta}\oint  \partial^{\alpha}
\big( B_x d y \wedge dz + B_y d z \wedge dx + B_z d x \wedge dy\big)
=  \frac{1}{2} Q m \theta_{\alpha\beta}v^{\beta}\oint  \partial^{\alpha} \vec{B} \cdot d\vec{S}\,,
\ene
and
\bee
\widetilde{\phi}_{\theta-v}^{AB-E} 
=  \frac{1}{2} Q m \theta_{\alpha\beta}v^{\beta}\oint  \partial^{\alpha}
\big( E_x dx + E_y dy + E_z d z\big) \wedge dt
= - \frac{1}{2} Q m \theta_{\alpha\beta}v^{\beta}\oint  \partial^{\alpha} \vec{A} \cdot d\vec{x}
= - \widetilde{\phi}_{\theta-v}^{AB-M} \,.
\ene
In the derivation above, we have used the relation $\vec{E}(t) = - \partial_{t}\vec{A}(t)$, which vilids because of the gauge invariance. The results indicate an exact cancellation between the magnetic and electric phase shifts in the momentum-dependent noncommutative correction. It is worthy to point out that this cancellation happens even in the case that there is time-position noncommutativity, \ie, $\theta^{0i}\neq0$.

For the momentum-independent noncommutative correction, because the derivation in \eqref{ncabam} for the time-dependent electromagnetic fields is still valid, therefore, the form of the magnetic phase shift does not change in this case,
\bee\label{ncabamt}
\widetilde{\phi}_{\theta-g}^{AB-M}(t)
 =  \frac{Q}{2} \big[ \vec{B}(t)\cdot \vec{\theta} ~\big] \phi^{AB}(t) \,.
\ene
Here we have write down the time dependence explicitly.
%\bee
%\widetilde{\phi}_{\theta-g}^{AB} 
%= - \frac{1}{2} Q^2 \theta_{ij}\oint
%\big( F^{2i} \partial^{3} A^{j} d y \wedge dz + F^{3i} \partial^{1} A^{j} d z \wedge dx + F^{1i} \partial^{2} A^{j} d x \wedge dy\big)\,.
%\ene
%and the electric part is
%\bee
%\widetilde{\phi}_{\theta-g}^{AB} 
%= - \frac{1}{2} Q^2 \theta_{ij} \oint 
%\big( F^{xi} \partial^{t} A^{j} dx + F^{yi} \partial^{t} A^{j} dy + F^{zi} \partial^{t} A^{j} d z\big) \wedge dt\,
%\ene
For the electric phase shift, inserting the relation $\vec{E}(t) = - \partial_{t}\vec{A}(t)$ we have
\bee\label{ncabaet}
\widetilde{\phi}_{\theta-g}^{AB-E} 
= -\frac{Q^2}{2} \oint   \vec{\theta}\cdot (\vec{E}\times\vec{A}) d t
= \frac{Q^2}{2} \oint   \vec{\theta}\cdot (\partial_{t}\vec{A}\times\vec{A}) d t
= 0\,,
\ene
where we have used the general relation $\vec{A}\times\vec{A}=0$ in the second step. Therefore, for the momentum-independent noncommutative corrections on the time-dependent AB phase shift, there is no cancelation between the magnetic and electric components. Therefore, the net time-dependent phase shift is,
\bee\label{ncabamt}
\widetilde{\phi}_{NC}^{AB}(t)
 =  \frac{Q}{2} \big[ \vec{B}(t)\cdot \vec{\theta} ~\big] \phi^{AB}(t) \,.
\ene
Compared to the time-independent AB phase shift \refb{netncab}, there is no overwhelming background which exists in the time-independent AB effect on both commutative and noncommutative space. Therefore, it is easier to measure the spatial noncommutativity by using the time-dependent AB effect. It is well know that the interference pattern will have a shift of $N$ fringes when the phase shift equals $N$ times of the flux quanta, $\Phi_{0}=h/e=4.13\times10^{-15}J\cdot s\cdot C^{-1}$ which is a very small quantity. Therefore, it is expected the time-dependent AB effect is sensitive to the noncommutative parameter. In addation, the noncommutative correction is enlarged by the large magnetic flux through the surface enclosed by the trajectory of incident charged particles.

\section{Detecting spatial noncommutativity}\label{sec:exp}
In last section we have shown that there is a nontrivial noncommutative correction on the time-dependent AB phase, see \eqref{ncabamt}. In this section we discuss how the spatial non commutativity can be probed by using the time-dependent AB effect, as well as the experimental sensitivity. 

In general the noncommutative correction \refb{ncabamt} oscillates with respect to time. Here we use the maximum value to illustrate the method of probing the spatial noncommutativity, and estimate the experimental sensitivity to the noncommutative parameter. {\em{Because there is no phase shift on the commutative space, therefore, the observation of the phase shift means there is spatial noncommutativity}}. Suppose there is a shift of $N$ fringes, then the noncommutative parameter can be determined by the \eqref{ncabamt} (with charge $\big|Q\big|=1$),
\bee\label{eq:theta}
\big|\theta\big|^{-1/2} 
%= 3.37\times 10^{-2} \frac{B\sqrt{S}}{\sqrt{N}} ~ {\rm GeV}
= 1.06\times 10^{-1} \sqrt{\frac{B\Phi}{N}} ~ {\rm GeV}\,,
\ene 
where $B$ is the magnetic field in unit of tesla ($\rm T$), and $\Phi$ is the magnetic flux in unit of $\rm T\cdot m^2$. In order to investigate the nontrivial relation implied by \eqref{eq:theta}, we introduce a variable $\overline{\Phi}$ defined by
\bee\label{eq:averageFlux}
\overline{\Phi} = \frac{\Phi}{N}\,.
\ene
Then, the magnetic field $B$ and the averaged magnetic flux $\overline{\Phi}$ are related by following relation,
\bee
\log(B)
%= 89.0\times \frac{1}{\big|\theta\big|\overline{\Phi} } ~ {\rm T}\,.
= 1.95 - \log(\big|\theta\big|) - \log(\overline{\Phi})\,.
\ene
Therefore, the logarithms of magnetic field and averaged magnetic flux possess an anti-collinear relation. It should be noticed that the \eqref{eq:averageFlux} is not well defined for the AB effect on the commutative space, because $N=0$ in this case. Therefore, we chose $\overline{\Phi}=\Phi$ to define this boundary. Then the magnetic field $B$ and the averaged flux $\overline{\Phi}$ have a collinear relation, \ie, $\log(B) \propto \log(\overline{\Phi})$, which is completely different from the behavior on noncommutative space. Therefore, by examining the relation between $\log(B)$ and $\log(\overline{\Phi})$, we can qualitatively test the spatial noncommutativity: {\it if it is collinear then the space is commutative, if it is anti-collinear then spatial noncommutativity exists}.

\draftnote{The bounds of the noncommutative parameter $\theta$ have been studied in various system. In Ref.~\cite{Chaichian:2002}, the authors studied the noncommutative corrections on the time-independent AB phase by using the path integral method. Even though a gauge invariant result was obtained, but the constraint was very week, a lower bound $\sqrt{\theta^{-1}}>10^{-6}\rm{GeV}$ was obtained. So far, the strongest bound comes from the clock-comparison tests, $\sqrt{\theta^{-1}} >10\rm{TeV}$ that was reported in Ref.~\cite{Carroll:2011}. However, the clock-comparison tests are related to the hyperfine splittings of atoms, and hence involves the spin degree of freedom which is strongly affected in expectation by the noncommutative algebra. Nevertheless, in general it is possible that the coupling parameters in the charge and spin degree of freedoms can have very different strengths~\cite{Carroll:2011,Colladay:1997,Colladay:1998}. Therefore, this strongest bound can be released for the quantum system in which spin does not play a role.

To measure the noncommutative correction on the time-dependent AB phase effect, it is important to point out the experimental conditions that have to be satisfied. On the ordinary space, to observe the cancellation of the magnetic and electric contributions, the oscillating rate of the external field have to be faster then the time scale that the charged particles get through the whole interference region~\cite{Singleton:2013}. This condition have to be satisfied for the noncommutative effect. In Refs.~\cite{Singleton:2015}, this condition was studied in detail by assuming that the length scale travelled by the charged particles is about $100{\rm\mu m}$. Even though so far, the time-dependent AB effect on ordinary space has not been observed experimentally, however it is expected to be observed in the near future. Below, we give an estimation of the experimental sensitivity on the noncommutative parameter $\theta$ based on the experimental achievements in the atom interference technology\cite{Cronin:2009}.
%However, the cancellation of the magnetic and electric contributions have been proved theoretically to be true for both Abelian~\cite{Rousseaux:2008,Moulopoulos:2010,Singleton:2013,Singleton:2015} and non-Abelian AB effect~\cite{Mansoori:2016}.
}

\begin{figure}[h]
\begin{center}
\includegraphics[scale=1.1]{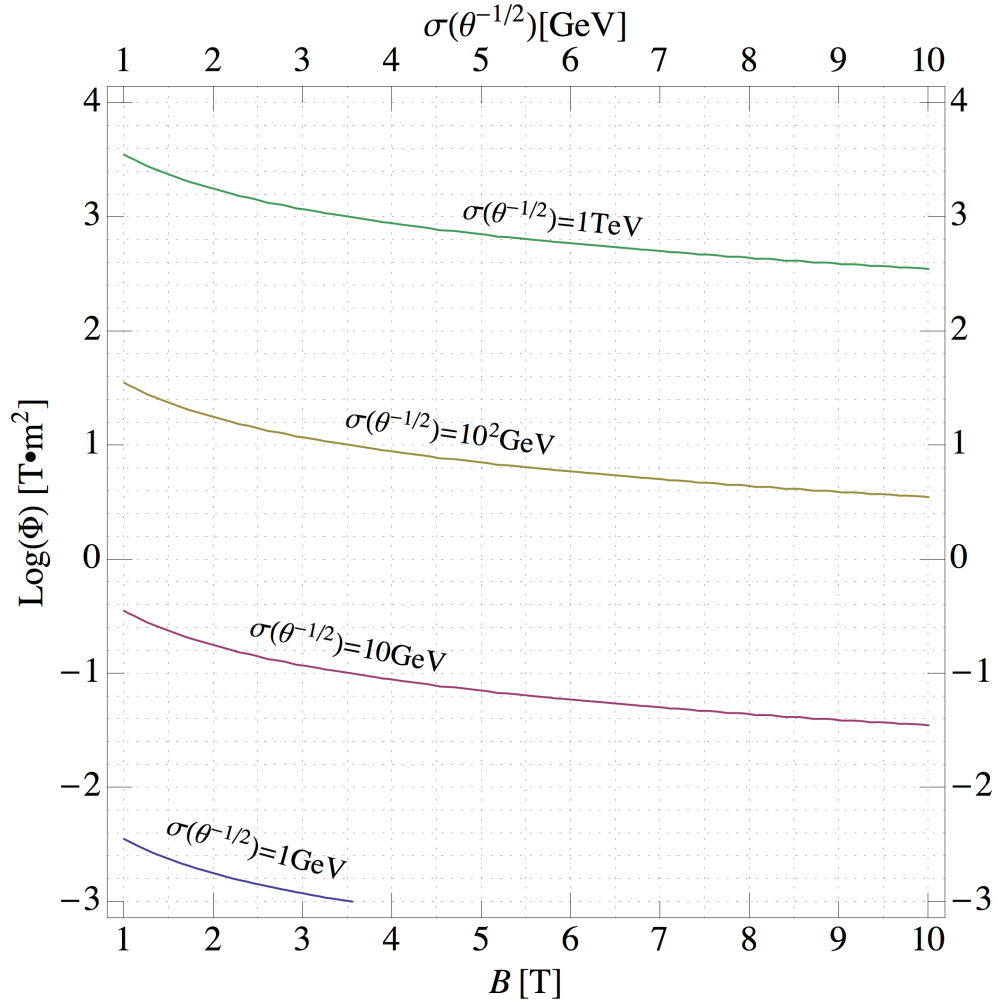}
\caption{Contour lines of the expected experimental sensitivity on the noncommutative parameter in the  $B-\log(\Phi)$ plane for $N=1$.}
\label{fig:sigmaTheta}
\end{center}
\end{figure}

The experimental sensitivity to the noncommutative parameter is estimated by using \eqref{eq:theta}.
The experimental precision for measuring the phase shift is about $0.1\%$~\cite{Cronin:2009}. Therefore, the sensitivity to the shift of fringe is about $0.0159\%$. For experiment with $B\overline{\Phi}$ at the order of $1$, and neglecting the uncertainties in the magnetic field $B$, the flux $\Phi$ and the number of shifts $N$, the sensitivity to the noncommutative parameter $\sigma(\big|\theta\big|^{-1/2})$ can reach $8.40{\rm GeV}$. This sensitivity can be enhanced by using stronger magnetic field strength, larger magnetic flux and higher experimental precision. Fig. \ref{fig:sigmaTheta} shows the contour lines of the sensitivity $\sigma(\big|\theta\big|^{-1/2})$ in the $B-\log(\Phi)$ plane for $N=1$. From it we can see that the experimental sensitivity is mainly affected by the magnetic flux $\Phi$. Because it is hard to enhance the magnetic field strength, therefore, increasing the length scale of the interference arms is essential to get better experimental sensitivity.

\section{Conclusions}\label{conclusion}
The AB effect is one of the most profound phenomena in the quantum gauge theory. In general there are two kinds of contributions: magnetic and electric phases, corresponding to the vector and scalar potentials, respectively. For the static case, \ie, time-independent AB effect, there is no local interactions between the charged particle and electromagnetic fields. However, this is not true for the time-dependent AB effect. It has been point out that, because of the electromagnetic gauge invariance and Lorentz invariance, the magnetic and electric contributions cancel each other in the time-dependent case, therefore, there is no net AB phase shift~\cite{Rousseaux:2008,Moulopoulos:2010,Singleton:2013,Singleton:2014,Singleton:2015}. Because of this, a tiny deviation from zero indicates new physics. Base on this observation, we studied the noncommutative corrections on the time-dependent AB effect.

To study the time-dependent AB effect on the noncommutative space, gauge invariant and Lorentz covariant formulations are necessary. We employ the SW map for this goal. Based on the SW map we studied both the time-independent and time-dependent AB effect. We find there are two kinds of noncommutative corrections in general: momentum-dependent and momentum-independent corrections. For the momentum-dependent noncommutative correction, there is a cancellation between the magnetic and electric phase shifts for both the time-independent and time-dependent AB effect, just like the case on the commutative space. However, there is important contribution in the momentum-independent noncommutative correction. This is also true for both the time-independent and time-dependent AB effect. The difference is that, for the time-dependent AB effect, there is no overwhelming background which exists in the time-independent AB effecton on both commutative and noncommutative space. Therefore, the time-dependent AB can be more sensitive to the spatial noncommutativity. 

The net noncommutative correction on the time-dependent AB phase shift is proportional to the magnetic flux through the fundamental area spanned by the noncommutative parameter $\theta$, \ie, $\vec{B}\cdot\vec{\theta}$. This correction is also scaled by the magnetic flux $\Phi$ through the surface enclosed by the trajectory of incident charged particles. An interesting result is that, the logarithms of magnetic field and averaged magnetic flux possess an anti-collinear relation. This is completely different from the collinear behavior on noncommutative space. Therefore, by examining the relation between $\log(B)$ and $\log(\overline{\Phi})$, we can qualitatively test the spatial noncommutativity. For $B\overline{\Phi}\sim 1$, our estimation on the experimental sensitivity shows that it can reach $8.4\rm GeV$. This sensitivity can be enhanced by using stronger magnetic field strength, larger magnetic flux, as well as higher experimental precision of the phase shift measuring. 

In summary we introduced a new approach to investigate the spatial noncmmutativity. The advance of this approach is the clean background because of vanishing net time-dependent AB phase on the commutative space. The experimental conditions that have to be satisfied in order to measure the noncommutative correction are also discussed.

\noindent\textbf{Acknowledgments}: 
K. M. is supported by the China Scholarship Council and the Hanjiang Scholar Project of Shaanxi University of Technology.  J. H. W. is supported by the National Natural Science Foundation of China under Grant No. 11147181 and the Scientific Research Project in Shaanxi Province under Grant No. 2009K01-54 and Grant No. 12JK0960. H.-X. Y. is supported in part by CNSF-10375052, the Startup Foundation of the University of Science and Technology of China and the Project of Knowledge Innovation Program (PKIP) of the Chinese Academy of Sciences.

%\section*{References}

\end{document}